\documentclass[12pt, a4paper] {article}

\usepackage{amsfonts}
\usepackage{verbatim}
\usepackage{color}

\begin{document}

\begin{center}

{\sf {\bf   Off-shell nilpotent (anti-)BRST symmetries for a free particle system on a toric geometry: superfield formalism}}

\vskip 2.5 cm

{\sf{ \bf R. Kumar}}\\
\vskip .1cm
{\it S. N. Bose National Centre for Basic Sciences,\\
Block JD, Sector III, Salt Lake, Kolkata$-$700098, India}\\
\vskip .1cm
{\sf E-mails: rohit.kumar@bose.res.in; raviphynuc@gmail.com}\\

\end{center}

\vskip 2.5 cm

\noindent

\noindent
{\bf Abstract:} We derive the  off-shell nilpotent as well as absolutely anticommuting (anti-)BRST symmetry transformations,  
within the framework of superfield approach to BRST formalism, for a free particle system constrained to move 
on a torus. We also construct the most appropriate gauge-fixed Lagrangian which respects the (anti-)BRST 
symmetry transformations.\\ 



\newpage

\section{Introduction}
 
The mathematical properties associated with the toric geometry have raised a great deal of interest of the 
theoretical physicists in the context of (super) string and gauge theories  \cite{green,pol}. For instance, 
the geometrical structure of a  closed string is one of the simplest examples of the torus. The Aharonov-Bohm 
and Casimir effects have also been studied in the realm of noncommutative toric geometry \cite{chai,mor}. 
The geometrical structure and physical characteristics  of the charged torus has been studied 
(see, e.g., \cite{nam1} for details).  Moreover, the square-root quantization of elementary particle masses 
and charges on the torus is also carried out \cite{nam2,nam3}.

The Becchi-Rouet-Stora-Tyutin (BRST) formalism is one of most elegant methods to quantize the gauge theories 
\cite{brs1,brs2,brs3,brs4}. In this formalism, the unitarity and the quantum ``gauge"  (i.e. BRST and anti-BRST) 
symmetries are respected together at any arbitrary order of perturbative computations. Recently, the BRST 
symmetries for a free particle system on a toric geometry have been discussed and its quantization has also 
been carried out in \cite{hong} in the context of Batalin-Fradkin-Vilkovisk formalism \cite{bfv,henn}.

The key properties (i.e. nilpotency and absolute anticommutativity) associated with the (anti-)BRST symmetry 
transformations have their geometric origin in the superfield formalism and these properties become transparent
\cite{b1,b2}. Within the framework of superfield approach to BRST formalism, the model of rigid rotor 
(see, e.g., \cite{nim}) has been studied where  a specific choice for the gauge potentials has been chosen to 
derive the proper (anti-) BRST symmetries\cite{malik1}. Furthermore, the model of rigid rotor provides a toy model 
for the Hodge theory where the continuous symmetries of the model provide the physical realizations of the de Rham 
cohomological operators of differential geometry and the discrete symmetry plays the role of Hodge duality operation 
\cite{malik1}. It is worthwhile to mention that, in the case of rigid rotor, the components of gauge potential 
transform in a completely different fashion as compared to other field-theoretic models (see, e.g., \cite{malik2,malik3,malik4,malik5}).        
Such study has led us to investigate the present model (i.e. a free particle on a torus) within 
the framework of superfield approach to  BRST formalism.

The contents of our present paper are organized as follows. In the next section, we briefly discuss about 
the free particle residing on a torus constrained to satisfy a geometrical constraint $(r - a) = 0$. 
We also discuss the Dirac brackets associated with the second-class constrained system. The third section is 
devoted to the conversion of the second-class constraints into  first-class constraints by the introduction 
of St{\"u}ckelberg like variables. The local gauge symmetries associated with the first-class constraints 
are also discussed in this section. The fourth section deals with the derivation of the off-shell nilpotent 
as well as absolutely anticommuting (anti-)BRST transformations within the framework of superfield formalism. 
In the fifth section, we construct an (anti-)BRST-invariant Lagrangian and derive the conserved (anti-)BRST charges. 
Finally, in the last section, we make some concluding remarks.

\section {Preliminaries: free particle system on toric geometry and Dirac brackets} 

We begin with a free particle of unit mass (i.e. $m = 1$) forced to satisfy a geometric constraint 
$(r -a) = 0$ on the torus with axial circle in the $x-y$ plane centered at origin of radius $l$, and circular 
cross-section of radius $r$. The Lagrangian for this constrained system can be written in terms of the 
toroidal coordinates ($r, \theta, \phi$) as (see, e.g., \cite{hong})
\begin{eqnarray}
L = \frac{1}{2}\, {\dot r}^2 + \frac{1}{2}\, r^2\,{\dot \theta}^2 
+ \frac{1}{2}\,\big(l + r \,sin\theta\big)^2 {\dot \phi}^2 + \lambda\,(r - a),
\end{eqnarray}
where $\dot r = d r/dt, \,\dot \theta = d \theta/dt,\, \dot \phi = d \phi/dt$ are the generalized velocities 
and $\lambda$ is a Lagrange multiplier. All the variables are function of the time evolution parameter $t$. 
The coordinates ($r, \theta, \phi$) for the toric geometry  
\begin{eqnarray}
 x= \big(l + r\, sin \theta\big) cos \phi, \quad y = \big(l + r\, sin \theta\big) sin \phi,\qquad 
 z =  r\, cos \theta,
\end{eqnarray}
satisfy the following relation: 
\begin{eqnarray}
r^2 = \big[\big(x^2 + y^2\big)^{1/2} - l\big]^2 + z^2,
\end{eqnarray}
where the angles $\theta$ and $\phi$ range from $0$ to $2\pi$.

The canonical  momenta $p_\lambda, \, p_r,\, p_\theta, \, p_\phi$ w.r.t. the dynamical variables
$\lambda,\, r, \, \theta, \, \phi$,  respectively,  are as follows
\begin{eqnarray}
p_\lambda = 0, \qquad p_r = \dot r, \qquad p_\theta = r^2\, \dot \theta,\qquad
 p_\phi = \big(l + r \, sin \theta\big)^2 \dot \phi. 
\end{eqnarray}
In the above, the vanishing momentum ($p_\lambda = 0$) is the primary constraint 
$\Lambda_1 = p_\lambda \approx 0$ on the theory \cite{dirac, sunder}.

By exploiting the Legendre transformations, the canonical Hamiltonian of the system can be written as
\begin{eqnarray}
H_c =\frac{p_r^2}{2} + \frac{p_\theta^2}{2 r^2} +\frac{p_\phi^2}{2 \big(l + r\,sin \theta\big)^2} - \lambda\, (r - a). 
\end{eqnarray}
In order to find the other constraints, we construct a primary Hamiltonian by adding the primary constraint
with an additional Lagrange multiplier  $\vartheta$ to the canonical Hamiltonian \cite{dirac,sunder}. Thus, 
the primary Hamiltonian $H_p$ (i.e. $ H_p= H_c + \vartheta \,p_\lambda$)  takes the following form
\begin{eqnarray}
H_p =\frac{p_r^2}{2} + \frac{p_\theta^2}{2 r^2} +\frac{p_\phi^2}{2 \big(l + r\,sin \theta\big)^2} 
- \lambda\, (r - a) + \vartheta \,p_\lambda.
\end{eqnarray}
According to Dirac,  constraints must remain intact with time \cite{dirac}. As a consequence, the time evolution 
of the primary constraint 
\begin{eqnarray}
\dot \Lambda_1 &=& \{\Lambda_1,\; H_p\}_{PB} = (r-a)\approx 0,
\end{eqnarray}
leads to the secondary constraint on the theory 
\begin{eqnarray}
\Lambda_2 &=& (r -a) \approx 0,
\end{eqnarray}
where in the computation of the above secondary constraint we have used the following non-vanishing 
Poisson brackets $\{~ ,~ \}_{PB}$, namely;
\begin{eqnarray}
&& \{r,\; p_r\}_{PB} = 1, \qquad \{\lambda,\; p_\lambda\}_{PB} = 1, \nonumber\\
&& \{\theta,\; p_\theta\}_{PB} = 1, \qquad \{\phi,\; p_\phi\}_{PB} = 1.  
\end{eqnarray}
Furthermore, the time evolution of the secondary constraint (i.e. $\dot \Lambda_2 = \{\Lambda_2,\; H_p\}_{PB} \approx 0$) 
yields the tertiary constraint 
\begin{eqnarray}
\Lambda_3 &=& p_r \approx 0.
\end{eqnarray}
Again, requiring $\Lambda_3$ to be time independent (i.e. $\dot \Lambda_3 \approx 0$), 
we have following quaternary constraint
\begin{eqnarray}
\Lambda_4 &=&  \frac{p_\theta^2}{r^3} + \frac{p_\phi^2\, sin \theta}{\big(l + r\, sin \theta\big)^3}
 + \lambda \approx 0.
\end{eqnarray}
Here, we point out that there are no further constraints because the time evolution of the 
quaternary constraint determines the value of Lagrange multiplier $\vartheta = 0$ and the series of 
constraints terminates here. Thus, we have, in totality,  four constraints on the theory, namely;
\begin{eqnarray}
&&\Lambda_1 = p_\lambda \approx 0, \qquad \Lambda_2 = (r -a) \approx 0, \qquad \Lambda_3 = p_r \approx 0, \nonumber\\
&& \Lambda_4 =  \frac{p_\theta^2}{r^3} + \frac{p_\phi^2\, sin \theta}{\big(l + r\, sin \theta\big)^3}
 + \lambda \approx 0.
\end{eqnarray}
We note that the above constraints form a set of second-class constraints in the language of  Dirac's 
classification scheme of the constraints \cite{dirac}.

Now we construct an antisymmetric $4\times4$ matrix $C_{ij}$  from the above constraints as follows
\begin{eqnarray}
C_{ij} = \{\Lambda_i, \;\Lambda_j\} =
  \left( {\begin{array}{cccc}
   0 & 0 & 0 & -1 \\
   0 & 0 & +1 & 0 \\
   0 & -1 & 0 & +\kappa\\
   +1 & 0 & - \kappa & 0\\
  \end{array} } \right).
\end{eqnarray}
where $i,j = 1, 2, 3, 4$ and $\kappa = 3\,p_\theta^2/r^4 + 3\, p_\phi^2 \, sin \theta/\big(l + r\, sin \theta\big)^4$.
It is evident from (13) that the matrix  $C_{ij}$ is a non-singular matrix whose determinant 
is $(+1)$. The inverse of the above matrix is  
\begin{eqnarray}
  C_{ij}^{-1}=
  \left( {\begin{array}{cccc}
   0 & +\kappa & 0 & +1 \\
   - \kappa & 0 & - 1 & 0 \\
   0 & +1 & 0 & 0\\
   -1 & 0 & 0 & 0\\
  \end{array} } \right).
\end{eqnarray}  
As a consequence, we can define the Dirac bracket. The Dirac bracket $\{F, G\}_D$ between any  two  dynamical 
variables $F(q,p)$ and $G(q, p)$ is given by   
\begin{eqnarray}
\{F,\, G\}_D = \{F,\, G\}_{PB} - \{F,\, \Lambda_i\}_{PB}\; C_{ij}^{-1}\; \{\Lambda_j, \, G\}_{PB}. \;\;
\end{eqnarray}
For our second-class system, the Dirac brackets among all the dynamical variables are 
\begin{eqnarray}
&& \{\theta,\; p_\theta\}_D = 1, \qquad \{\phi,\; p_\phi\}_D = 1, \qquad
 \{\lambda,\; \theta\}_D = \frac{2\, p_\theta}{r^3},  \nonumber\\
&& \{\lambda,\; \phi\}_D = \frac{2\, p_\phi\, sin \theta}{(l + r\, sin \theta)^3},\quad
 \{\lambda,\; p_\theta\}_D = - \frac{p_\phi^2\, (l - 2r\,sin \theta)\, cos \theta}{(l + r \, sin \theta)^4}. \qquad
\end{eqnarray}
All the rest of the  Dirac brackets turn out to be zero.

To quantize the above constrained  system, we promote all the Dirac's Poisson brackets 
(at the classical level) to the commutators (at the quantum level) and replace all the 
dynamical variables by their operator form \cite{dirac,sunder}.

\section{First-class constraints and gauge symmetry} 

It is evident from our earlier discussions  that the present system is endowed with a set of second-class constraints. 
To convert the second-class constraints into the first-class constraints, we redefine the variables:
 $r \to r- \xi$ and $ p_r \to p_r + p_\xi$ (see, e.g., \cite{hong}). The canonical Hamiltonian (5) takes the following form:
\begin{eqnarray}
{\tilde H}_c &=&\frac{(p_r + p_\xi)^2}{2} + \frac{p_\theta^2}{2\,(r - \xi)^2} 
+\frac{p_\phi^2}{2\,\big[l + (r - \xi)\,sin \theta\big]^2} - \lambda\, (r - \xi - a), \qquad
\end{eqnarray}
and the corresponding first-order Lagrangian is 
\begin{eqnarray}
L_f &=& \dot r p_r + \dot \theta  p_\theta + \dot \phi\, p_\phi + \dot \xi  p_\xi 
- \frac{(p_r + p_\xi)^2}{2} - \frac{p_\theta^2}{2(r - \xi)^2} \nonumber\\
&-& \frac{p_\phi^2}{2\big[l + (r - \xi)\,sin \theta\big]^2} + \lambda \,(r - \xi - a). 
\end{eqnarray}
The above Lagrangian (18) is endowed with the primary constraint $\Omega_1 = p_\lambda \approx 0$ and its  
time evolution leads to the secondary constraint $\Omega_2 = (r - \xi - a) \approx 0$. It is to be
noted that there are no further constraints because one can show that $\Omega_2$ commutes with the 
canonical Hamiltonian $\tilde H_c$. Both the constraints (i.e. $\Omega_1$ and $\Omega_2$) are first-class 
constraints in the Dirac's terminology and this is a signature of the gauge theory.

The most general form of the generator in terms of the first-class constraints, which generates the gauge 
transformations, can be written as
\begin{eqnarray}
Q = \dot \chi\, p_\lambda - \chi \,(r -\xi  -a),
\end{eqnarray}      
where $\chi = \chi(t)$ is an infinitesimal and local gauge parameter. The above generator generates the following 
gauge transformations for all  the dynamical variables, namely;
\begin{eqnarray}
\delta_{(gt)} \lambda = \dot \chi, \quad \delta_{(gt)} p_r = \chi, \quad \delta_{(gt)} p_\xi = - \chi,\quad
\delta_{(gt)} [r,\, \theta,\, \phi,\, \xi, \,p_\theta,\, p_\phi] = 0. 
\end{eqnarray} 
Under the above gauge transformations, the  first-order Lagrangian (18)  transforms to a total time derivative:
\begin{eqnarray}
\delta_{(gt)}\, L_f = \frac{d}{dt}\big[\chi \big(r - \xi - a\big)\big].
\end{eqnarray}
Thus, the action integral (i.e. $S = \int dt \, L_f$) remains invariant under the above gauge transformations (20).

\section {Off-shell nilpotent (anti-)BRST symmetries: superfield formalism}

It is clear from our earlier section, the existence of the first-class constraints ($\Omega_ 1 = p_\lambda \approx 0$ 
and $\Omega_2 = (r - \xi - a) \approx 0$) implies that the modified theory is a gauge theory with the gauge 
potentials $\lambda, \, p_r$ and $p_\xi$. These gauge potentials transform in a completely different way under 
the gauge transformations (20). As a consequence, we define the exterior derivative $d$ 
and the 1-form connection, respectively    
\begin{eqnarray}
d &=& dt\, \partial_t + dr\, \partial_r + d\xi\, \partial_\xi, \qquad d^2 =0, \nonumber\\
A^{(1)} &=& dt\, \lambda(r, \xi, t) + dr\, B(r, \xi, t) - d\xi\, E(r, \xi, t), \quad
\end{eqnarray}
such that our ordinary 3D space is characterized by three coordinates ($r, \xi, t$). 
In the superfield formalism  \cite{b1,b2}, we assume that these coordinates are independent variables. We shall 
see, later on, that the gauge potential components $B(r, \xi, t)$ and $E(r, \xi, t)$ would be connected with 
$p_r$ and $p_\xi$, respectively. Finally, at the end of computations, we shall take the limit $(r,\, \xi) \to 0$ 
so that all the dynamical variables of the present theory become only the function of time evolution parameter $t$.   
Furthermore, we have taken the negative sign in the third term of the 1-form connection $A^{(1)}$ so that the 
2-form curvature $dA^{(1)}$ would remain invariant under the gauge and/or (anti-)BRST symmetry transformations. 
The 2-form curvature is given by
\begin{eqnarray}
d A^{(1)}&=& (dt \wedge dr)\,\big[\partial_t\, B(r, \xi, t)- \partial_r \, \lambda(r, \xi, t)\big]\nonumber\\
&-& (dt \wedge d\xi)\,\big[\partial_t\, E(r, \xi, t) + \partial_\xi \, \lambda(r, \xi, t)\big]\nonumber\\
&-& (dr \wedge d\xi)\,\big[\partial_r\, E(r, \xi, t)+ \partial_\xi \, B(r, \xi, t)\big],\quad
\end{eqnarray}
where we have used the following properties of the wedge products: $dt \wedge dt = 0$, $dr \wedge dr = 0$, 
$d\xi \wedge d\xi = 0$, $dt \wedge dr = - dr \wedge dt$, $dt \wedge d\xi = - d\xi \wedge dt$, $dr \wedge d\xi = - d\xi \wedge dr.$

In the superfield formalism, we generalize our 3D space to (3, 2)D superspace. The (3, 2)D 
superspace is parametrized, in addition to the ordinary bosonic variables $(r, \xi, t)$, by a pair of Grassmannian variables  
$(\eta, \bar \eta)$  (with $\eta^2 = \bar\eta^2 = 0,\; \eta\bar\eta + \bar\eta \eta = 0$). The exterior 
derivative $d$ and the 1-form connection $A^{(1)}$ are also generalized to the super exterior derivative $\tilde d$ 
(with ${\tilde d}^2 = 0$)  and super 1-form connection $\tilde A^{(1)}$ onto (3, 2)D supermanifold as  
\begin{eqnarray}
d \rightarrow \tilde d &=& dt\, \partial_t + dr\, \partial_r + d\xi\, \partial_\xi + d\eta\, \partial_\eta 
+ d\bar\eta\, \partial_{\bar\eta}, \nonumber\\
A^{(1)}\rightarrow  \tilde A^{(1)} &=& dt\, \tilde \lambda(r, \xi, t, \eta, \bar \eta) 
+ dr\, \tilde B(r, \xi, t, \eta, \bar \eta)- d\xi\, \tilde E(r, \xi, t, \eta, \bar \eta)\nonumber\\
& +& d\eta\, \bar F(r, \xi, t, \eta, \bar \eta) + d\bar \eta\, F(r, \xi, t, \eta, \bar \eta), 
\end{eqnarray}
where $\partial_\eta = \partial/\partial \eta$ and $\partial_{\bar \eta} = \partial/\partial \bar \eta$ are 
the Grassmannian translational generators along 
 $\eta$, $\bar \eta$ directions, respectively.  The components of the supervariables can be expanded 
along the Grassmannian directions as follows   
\begin{eqnarray}
\tilde \lambda(r, \xi, t, \eta, \bar \eta) &=& \lambda(r, \xi, t) + \eta\, \bar f_1(r, \xi, t) 
+ \bar \eta \, f_1(r, \xi,t) + i\,\eta\,\bar\eta \,B_1(r, \xi, t),\nonumber\\
\tilde B(r, \xi, t, \eta, \bar \eta) &=& B(r, \xi, t) + \eta\, \bar f_2(r, \xi, t)
+ \bar \eta \, f_2(r, \xi,t) + i\,\eta\,\bar\eta \,B_2(r, \xi, t),\nonumber\\
\tilde E(r, \xi, t, \eta, \bar \eta) &=& E(r, \xi, t) + \eta\, \bar f_3(r, \xi, t)
+ \bar \eta \, f_3(r, \xi,t) + i\,\eta\,\bar\eta \,B_3(r, \xi, t),\nonumber\\
F(r, \xi, t, \eta, \bar \eta) &=& C(r, \xi, t) + i\,\eta\, \bar b_1(r, \xi, t)
+ i\,\bar \eta \, b_1(r, \xi,t) + i\,\eta\,\bar\eta \,s(r, \xi, t),\nonumber\\
\bar F(r, \xi, t, \eta, \bar \eta) &=& \bar C(r, \xi, t) + i\,\eta\, \bar b_2(r, \xi, t)
+ i\,\bar \eta \, b_2(r, \xi,t) + i\,\eta\,\bar\eta \,\bar s(r, \xi, t). \qquad
\end{eqnarray}
Here $b_1$, $\bar b_1$,  $b_2$,  $\bar b_2$, $B_1$, $B_2$, $B_3$ are the secondary bosonic fields and
the secondary fields $f_1$, $\bar f_1$, $f_2$, $\bar f_2$, $f_3$, $\bar f_3$, $s$, $\bar s$ are  fermionic 
in nature. The (anti-) ghost fields $(\bar C)C$ (with $C^2 = \bar C^2 = 0, \; C\, \bar C + \bar C\, C =0$) are  
also fermionic in nature. These secondary (bosonic) fermionic 
fields can be determined, with the help of horizontality condition (HC), in terms of the basic 
and auxiliary fields of the present theory. The following mathematical form of HC 
\begin{eqnarray}
\tilde d\, \tilde A^{(1)} = d\, A^{(1)}, 
\end{eqnarray}
implies that the l.h.s. has to be independent of the Grassmannian variables $(\eta, \bar\eta)$ 
when the $d A^{(1)}$ is generalized onto the (3, 2)D supermanifold. The explicit form of the l.h.s. is   
\begin{eqnarray}
\tilde d \tilde A^{(1)}&=& (dt \wedge dr)\big[\partial_t \tilde B - \partial_r  \tilde  \lambda\big]
- (dt \wedge d\xi)\big[\partial_t \tilde  E + \partial_\xi  \tilde  \lambda \big]\nonumber\\
&-& (dr \wedge d\xi)\big[\partial_r \tilde  E + \partial_\xi  \tilde  B \big] 
+ (dt \wedge d\eta)\big[\partial_t  \bar F - \partial_\eta  \tilde \lambda \big]\nonumber\\
&+& (dt \wedge d \bar \eta)\big[\partial_t F - \partial_{\bar \eta}  \tilde  \lambda \big] 
+ (dr \wedge d\eta)\big[\partial_r  \bar F - \partial_\eta  \tilde B \big] \nonumber\\
&+& (dr \wedge d \bar \eta)\big[\partial_r F - \partial_{\bar \eta}  \tilde  B \big] + (d\xi \wedge d\eta)\big[\partial_\xi  \bar F + \partial_\eta  \tilde E \big] \nonumber\\
&+&(d \xi \wedge d \bar \eta)\big[\partial_\xi F + \partial_{\bar \eta}  \tilde  E \big] 
-(d \eta \wedge d \bar \eta)\big[\partial_\eta F + \partial_{\bar \eta}  \bar F \big] \nonumber\\
&-& (d \bar \eta \wedge d \bar \eta)\, \partial_{\bar \eta} F - (d \eta \wedge d\eta)\, \partial_\eta \bar F . 
\end{eqnarray}
Exploiting HC, we obtain the following algebraic relationships among the secondary and basic fields, namely; 
\begin{eqnarray}
&& b_1 = 0,\; \quad \bar b_2 = 0, \;\quad s = 0, \;\quad \bar s = 0, \;\quad  f_1 = \dot C, \;\quad
 \bar f_1  = \dot {\bar C}, \;\quad f_2 = \partial_r C, \nonumber\\
&& \bar f_2 = \partial_r \bar C, \qquad \bar b_1 + b_2 = 0, \qquad
 B_1 = \dot b_2 = - \dot{\bar b}_1, \qquad B_2 = \partial_r b_2 = - \partial_r \bar b_1,\nonumber\\
&& f_3 = - \partial_\xi C, \qquad
\bar f_3 = - \partial_\xi \bar C, \qquad B_3 = - \partial_\xi b_2 = \partial_\xi \bar b_1, \qquad
 \dot B_3 = - \partial_\xi B_1, \nonumber\\
&& \partial_r B_3 = - \partial_\xi B_2, \quad \dot f_2 = \partial_r f_1, \;\quad
\dot {\bar{ f_2}} = \partial_r \bar f_1, \quad \dot f_3 = - \partial_\xi f_1, \;\quad \dot B_2 = \partial_r B_1, \nonumber\\
&& \dot {\bar {f_3}} = - \partial_\xi \bar f_1, \qquad
  \partial_r f_3 = - \partial_\eta f_2, \qquad \partial_r \bar f_3 = - \partial_\xi \bar f_2.
\end{eqnarray}
In the above, we make the choice $b_2 = - \bar b_1 = b$ and we get the following expressions for the supervariables
\begin{eqnarray}
\tilde \lambda^{(R)}(r, \xi, t, \eta, \bar \eta) &=& \lambda(r, \xi, t) + \eta\, \dot {\bar C}(r, \xi, t)
+ \bar \eta \, \dot C(r, \xi,t) + i\,\eta\,\bar\eta \, \dot b(r, \xi, t),\nonumber\\
\tilde B^{(R)}(r, \xi, t, \eta, \bar \eta) &=& B(r, \xi, t) + \eta\, \partial_r \bar C(r, \xi, t)
+ \bar \eta \,  \partial_r C(r, \xi,t) \nonumber\\
&+& i\,\eta\,\bar\eta \, \partial_r b(r, \xi, t),\nonumber\\
\tilde E^{(R)}(r, \xi, t, \eta, \bar \eta) &=& E(r, \xi, t) - \eta\,  \partial_\xi \bar C(r, \xi, t) 
- \bar \eta\,\partial_\xi C(r, \xi,t) \nonumber\\
&-& i\,\eta\,\bar\eta \, \partial_\xi b(r, \xi, t),\nonumber\\
F^{(R)}(r, \xi, t, \eta, \bar \eta) &=& C(r, \xi, t) - i\,\eta\, b(r, \xi, t),\nonumber\\
\bar F^{(R)}(r, \xi, t, \eta, \bar \eta) &=& \bar C(r, \xi, t) + i\,\bar \eta \, b(r, \xi,t),
\end{eqnarray}
where all the variables on the r.h.s. are the functions of ($r, \xi,t$). The superscript $(R)$ denotes the 
reduced form of the super expansions.

At this juncture, we make  judicious choices for the gauge potentials $B(r, \xi, t)$ and 
$E(r, \xi, t)$ in terms of the gauge components $p_r(r, \xi, t)$ and $p_\xi (r,\xi,t)$, respectively,
and for their generalizations in the superspace, too. These choices are as follows   
\begin{eqnarray}
 B(r, \xi, t) = \partial_r p_r(r, \xi, t), \qquad \tilde B^{(R)}(r, \xi, t, \eta, \bar \eta) = \partial_r \tilde P_r^{(R)}(r, \xi, t, \eta, \bar \eta), \nonumber\\ 
 E(r, \xi, t) = \partial_\xi p_\xi(r, \xi, t), \qquad  \tilde E^{(R)}(r, \xi, t, \eta, \bar \eta) = \partial_\xi \tilde P_\xi^{(R)}(r, \xi, t, \eta, \bar \eta).
\end{eqnarray}
Exploiting (29) and (30), we obtain
\begin{eqnarray}
\tilde P_r^{(R)}(r, \xi, t, \eta, \bar \eta) =  p_r(r, \xi, t) + \eta \, \bar C(r, \xi, t) + \bar \eta C(r, \xi, t) + i\,\eta\,\bar\eta\,b (r, \xi, t), \nonumber\\ 
\tilde P_\xi^{(R)}(r, \xi, t, \eta, \bar \eta) = p_\xi(r, \xi, t) - \eta \, \bar C (r, \xi, t) - \bar \eta C (r, \xi, t) - i\,\eta\,\bar\eta\,b(r, \xi, t).
\end{eqnarray}
Taking the limit $r \to 0$ and $\xi \to 0$ in equations (29) and (31), we obtain the physical super expansions 
on the (1, 2)D supermanifold. These super expansions are 
\begin{eqnarray}
\tilde \lambda^{(h)}(t, \eta, \bar \eta) &=& \lambda(t) + \eta\, \dot {\bar C}(t) + \bar \eta \, 
\dot C(t) + i\,\eta\,\bar\eta \, \dot b(t) \nonumber\\
&\equiv&  \lambda(t) + \eta\, (s_{ab}\, \lambda) + \bar \eta \, (s_b\, \lambda) + \eta\,\bar\eta \, (s_b\,s_{ab}\, \lambda),\nonumber\\
\tilde P_r^{(h)}(t, \eta, \bar \eta) &=&  p_r(t) + \eta \, \bar C(t) + \bar \eta \,C(t) + i\,\eta\,\bar\eta\,b(t) \nonumber\\
&\equiv& p_r(t) + \eta\, (s_{ab}\, p_r) + \bar \eta \, (s_b\, p_r) + \eta\,\bar\eta \, (s_b\,s_{ab}\, p_r),\nonumber\\
\tilde P_\xi^{(h)}(t, \eta, \bar \eta) &=& p_\xi(t) - \eta \, \bar C(t) - \bar \eta \,C(t) - i\,\eta\,\bar\eta\,b(t) \nonumber\\
&\equiv& p_\xi(t) + \eta\, (s_{ab}\, p_\xi) + \bar \eta \, (s_b\, p_\xi) + \eta\,\bar\eta \, (s_b\,s_{ab}\, p_\xi),\nonumber\\
F^{(h)}(t, \eta, \bar \eta) &=& C(t) - i\,\eta\, b(t) \nonumber\\
& \equiv& C(t) + \eta\, (s_{ab}\, C) + \bar \eta \, (s_b\, C) + \eta\,\bar\eta \, (s_b\,s_{ab}\, C),\nonumber\\
\bar F^{(h)}(t, \eta, \bar \eta) &=& \bar C(t) + i\,\bar \eta \, b(t)\nonumber\\
& \equiv& \bar C(t) + \eta\, (s_{ab}\, \bar C) + \bar \eta \, (s_b\, \bar C) + \eta\,\bar\eta \, (s_b\,s_{ab}\, \bar C),
\end{eqnarray}
where the superscript ($h$) on the l.h.s. denotes the super expansions obtained after
the application of HC.

We point out that the following  dynamical  variables $r$, $\theta$, $p_\theta$, $p_\phi$ remain invariant 
under the gauge transformations [cf. (20)]. Thus, we can apply the ``augmented" superfield formalism 
\cite{malik6,malik7,malik8,rohit} which demands that the gauge-invariant (physical) quantities remain independent 
of the Grassmannian variables. As a consequence,  the supervariables  corresponding to the  (ordinary)
variables $r$, $\theta$, $p_\theta$, $p_\phi$ remain unaffected by the presence of Grassmannian variables 
when the latter variables  are generalized onto the (1, 2)D supermanifold. Mathematically, this statement 
can be corroborated as 
\begin{eqnarray}
&&\tilde r (t, \eta, \bar\eta) = r (t), \qquad \tilde \xi (t, \eta, \bar\eta) = \xi (t),\nonumber\\
&& {\tilde \theta} (t, \eta, \bar\eta) =  \theta (t), \qquad \tilde P_\theta (t, \eta, \bar\eta) = p_\theta (t),  \nonumber\\
&&{\tilde \phi} (t, \eta, \bar\eta) =  \phi(t),  \qquad  \tilde P_\phi (t, \eta, \bar\eta) = p_\phi (t). 
\end{eqnarray}
From the above super expansions [cf. (32) and (33)], one can easily read out the (anti-)BRST transformations
for all the variables. For instance, the BRST transformation ($s_b$) is equal to the translation of the superfield 
along  $\bar\eta$-direction while keeping  $\eta$-direction fixed. Similarly, the anti-BRST transformation 
($s_{ab}$) can be obtained by taking the translation of the superfield along $\eta$-direction and $\bar \eta$-direction remains 
intact. The above statements can be, mathematically, expressed as
\begin{eqnarray}
s_b \Phi (t) &=& \frac{\partial}{\partial \bar \eta} \tilde \Phi (t, \eta, \bar \eta)\Big|_{\eta = 0}, \nonumber\\
s_{ab} \Phi (t) &=& \frac{\partial}{\partial \eta} \tilde \Phi (t, \eta, \bar \eta)\Big|_{ \bar\eta = 0}, \nonumber\\
s_b s_{ab}\Phi (t) &=& \frac{\partial}{\partial \bar \eta} \frac{\partial}{\partial \eta} \tilde \Phi (t, \eta, \bar \eta),
\end{eqnarray}  
where $\tilde \Phi (t, \eta, \bar \eta)$ is the superfield corresponding to the generic dynamical variable 
$\Phi(t) \equiv r, \theta, \phi, \xi, \lambda, p_r, p_\theta, p_\phi,$ $p_\xi$, $C, \bar C$. Exploiting the first two equations 
in (34), we obtain the following  off-shell nilpotent (i.e. $s^2_{(a)b} = 0$)  as well as absolutely 
anticommuting (i.e. $s_b\, s_{ab} + s_{ab}\, s_b  = 0$) (anti-)BRST symmetry  transformations ($s_{(a)b}$):
\begin{eqnarray}
&&s_{ab} \, \lambda = \dot {\bar C}, \qquad s_{ab}\, p_r = \bar C, \qquad s_{ab}\, p_\xi = - \bar C, \qquad 
s_{ab}\, C = - i\,b, \nonumber\\
&&  s_{ab}\,[r,\,\theta,\, \phi,\, \xi,\, b,\, \bar C,\, p_\theta,\, p_\phi],
\end{eqnarray}
\begin{eqnarray}
&& s_b \, \lambda = \dot C, \qquad s_b\, p_r = C, \qquad s_b\, p_\xi = - C, \qquad 
s_b\, \bar C = i\,b, \qquad \nonumber\\
&& s_b\,[r,\,\theta,\, \phi,\, \xi,\, b,\, C,\, p_\theta,\, p_\phi].
\end{eqnarray}
Furthermore, the (anti-)BRST transformations of the auxiliary field $b$ have been derived 
from the requirements of the nilpotency and/or absolute anticommutativity properties of the (anti-)BRST 
transformations. We point out that, in addition to (35) and (36), the last equation in (34) implies the 
following non-vanishing transformations: $s_b\,s_{ab}\, \lambda = i\, \dot b, \; s_b\,s_{ab}\, 
p_r = i\,b,\; s_b\,s_{ab}\,p_\xi = - i\, b$.

\section{(Anti-)BRST-invariant  Lagrangian and conserved charges}

The  gauge-fixed Lagrangian which respects the off-shell nilpotent 
(anti-)BRST symmetry transformations can be written as 
\begin{eqnarray}
L_b &=& L_f - s_b \left[i\, \bar C \left(\dot \lambda - p_r + p_\xi + \frac{b}{2}\right) \right]\nonumber\\ 
&\equiv& L_f + s_{ab} \left[i\, C \left(\dot \lambda - p_r + p_\xi + \frac{b}{2}\right) \right],
\end{eqnarray}
where $L_f$ is our previous Lagrangian [cf. (18)].
The Lagrangian $L_b$ in its full blaze of glory can be written as
\begin{eqnarray}
L_b = L_f + \frac{b^2}{2}+ b\,\big(\dot \lambda - p_r + p_\xi\big) - i\, \dot{\bar C}\, \dot C - 2i\, \bar C \,C.
\end{eqnarray}
Here $b$ is the Nakanishi-Lautrup type auxiliary variable. One can checked that the continuous (anti-)BRST 
transformations leave the Lagrangian (38) quasi-invariant. To be more precise, under the (anti-)BRST 
transformations, the Lagrangian transforms to a total time derivative 
\begin{eqnarray}
s_b\,L_b &=& \frac{d}{dt}\big[C\big(r - \xi - a\big) + b\, \dot C\big],\nonumber\\ 
s_{ab}\,L_b &=& \frac{d}{dt}\big[\bar C\big(r - \xi - a\big) + b\, \dot {\bar C}\big].
\end{eqnarray}
Thus, the action integral (i.e. $S = \int dt\,L_b$) remains invariant under the 
(anti-)BRST transformations $(s_{(a)b})$. According to  Noether's theorem, 
the invariance of the action under  the above continuous 
(anti-)BRST symmetry transformations leads to the following conserved (i.e. $\dot Q_{(a)b}$)
(anti-)BRST charges $Q_{(a)b}$, namely;
\begin{eqnarray}
Q_{ab} = b\, \dot {\bar C} - \dot b\, \bar C, \qquad Q_b = b\, \dot C - \dot b\, C.
\end{eqnarray}
The conservation law ($\dot Q_{(a)b} = 0$) can be proven by exploiting the 
following Euler-Lagrange equations of motion:
\begin{eqnarray}
&& \dot b = (r- \xi - a), \qquad b = - (\dot \lambda - p_r + p_\xi), \qquad \dot \theta = \frac{p_\theta}{(r - \xi)^2}, \nonumber\\
&&\dot p_r = \frac{p_\theta^2}{(r - \xi)^3} + \frac{p_\phi^2}{[l + (r - \xi)sin \theta]^3} - \lambda, 
\qquad \lambda = \dot p_\xi, \nonumber\\
&& \dot r = (p_r + p_\xi) + b, \qquad \dot \xi = (p_r + p_\xi) - b,  \qquad \dot p_\phi = 0, \nonumber\\
&& \dot \phi = \frac{p_\phi}{[l + (r - \xi)sin \theta]^2}, \qquad
\dot p_\theta = \frac{p_\theta^2 (r - \xi)cos \theta}{[l + (r - \xi)sin \theta]^3},\nonumber\\
&&\ddot{\bar C} -2\bar C = 0, \qquad \ddot{C} -2 C = 0. 
\end{eqnarray}
The above equations of motion have been derived from the Lagrangian (38).

We point out that the nilpotency (i.e. $Q_{(a)b}^2 = 0$) and the anticommutativity 
(i.e. $Q_b\,Q_{ab} + Q_{ab}\,Q_b = 0$) of the (anti-)BRST charges
$Q_{(a)b}$ can be proven in a simple and straightforward manner 
\begin{eqnarray}
 s_b Q_b &=& -i \{Q_b, Q_b\} = 0\; \Rightarrow \;Q_b^2 =0, \qquad\qquad\; \qquad\nonumber\\
 s_{ab} Q_{ab} &=& -i \{Q_{ab}, Q_{ab}\} = 0 \;\Rightarrow \;Q_{ab}^2 =0, \qquad \;\qquad\nonumber\\
 s_b Q_{ab} &=& -i \{Q_{ab}, Q_b\} = 0 \;\Rightarrow\; Q_b Q_{ab} + Q_{ab} Q_b=0,\nonumber\\
 s_{ab} Q_b &=& -i \{Q_b, Q_{ab}\} = 0 \;\Rightarrow \;Q_b Q_{ab} + Q_{ab} Q_b=0. 
\end{eqnarray} 
In the above, the first and second  lines show the nilpotency of the BRST and anti-BRST charges, respectively
whereas the third and fourth lines imply the absolute anticommutativity of the (anti-)BRST charges.

Before we close this section, we lay emphasis on the fact that the physicality condition 
$Q_{(a)b}|phys\rangle = 0$, on the conserved (anti-)BRST charges $Q_{(a)b}$, yields 
\begin{eqnarray}
b\,|phys\rangle &=& 0\; \Rightarrow\;  p_\lambda|phys\rangle = 0,\nonumber\\
\dot b\,|phys\rangle &=& 0 \; \Rightarrow\;  (r - \xi - a)\,|phys\rangle = 0,
\end{eqnarray}
where $p_\lambda = \partial L_b/\partial \dot \lambda = b$ is the canonical momentum w.r.t the dynamical variable  $\lambda$. 
These conditions imply that the operator form of the first-class constraints $\Omega_1 = p_\lambda \approx  0, \; 
\Omega_2 = (r - \xi -a) \approx 0$ annihilates the physical state $(|phys\rangle)$ of the total quantum Hilbert space of states.  
As a result, the above physicality criteria $Q_{(a)b}|phys\rangle = 0$ are consistent with the Dirac's quantization of the
constrained system \cite{dirac,sunder}.

\section{Conclusions}
In our present investigation, we have studied a free particle residing on a torus  constrained to satisfy the 
geometric constraint $(r -a)= 0$. This model is endowed with a set of four second-class constraints in the 
Dirac terminology. Thus,  we have derived the Dirac brackets. Furthermore, by incorporating the St{\"u}ckelberg 
like variables $(\xi, p_\xi)$, the second-class constraints turn into the first-class constraints. As a result, 
the modified theory respects the gauge symmetries.  These gauge transformations have been derived from the 
first-class constraints.

We lay emphasis on the fact that the components of gauge potentials (i.e. $\lambda, p_r, p_\xi$) transform quite 
differently under the gauge transformations (20). Thus, the derivation of the proper (anti-)BRST transformations 
was not straightforward. However, a similar kind of work has been carried out for the model of rigid rotor 
\cite{malik1} where a non-trivial choice for the gauge potential has been made which is quite different from the 
other gauge field theoretic models (see, e.g., \cite{malik2, malik3}). This study has led us to construct a 
1-form connection [cf. (22)] and we have made some judicious choices for the (super) gauge potentials [cf. (30)]. Within the 
framework of superfield formalism, we have incorporated the gauge components $p_r$ and $p_\xi$  very carefully [cf. (30)].     
Exploiting the celebrated HC and taking the limits $r\to 0,\; \xi \to 0$ (at the end of computations) in the 
super expansions (29) and (31), we  obtain the proper off-shell nilpotent as well as absolutely anticommuting 
(anti-)BRST transformation [cf. (35), and (36)].

It is to be noted that the variable $\lambda$ is no longer a Lagrange multiplier  within the framework of BRST formalism. As it 
can be checked, the (anti-) BRST-invariant Lagrangian (38) contains a time derivative of $\lambda$. Thus, it is a dynamical variable.
The continuous (anti-)BRST symmetry transformations lead to the conserved, nilpotent and anticommuting (anti-)BRST charges.     
Furthermore, the physicality criteria $Q_{(a)b}|phys\rangle = 0$ produce the first-class constraints 
$p_\lambda \approx 0, \; (r - \xi - a) \approx 0$ [cf. (43)]. Thus, the physicality criteria $Q_{(a)b}|phys\rangle = 0$
establish the connection between the BRST quantization and the Dirac's quantization of constrained system \cite{dirac,sunder}.

We point out that our present approach can also be generalized for the relativistic field-theoretic models
(see, e.g., \cite{malik4,gupta}) where the celebrated HC and  gauge-invariant restriction(s)  play an important role 
to derive the nilpotent (anti-)BRST symmetry transformations within the framework of superfield formalism.
It would be a nice work to look for the generalization of the present approach for the higher-form and higher dimensional
 (non-) Abelian gauge field theories.
It is interesting to point out that, within the framework of BRST formalism, the present model would turn out 
to be a model for the  Hodge theory where all the de Rham cohomological operators of differential geometry 
finds their physical realizations in terms of the symmetry properties \cite{malik1,malik2,malik3,malik4,malik5,gupta}. 
We shall investigate this issue in our future work \cite{rk}.


\end{document}